\documentclass[manuscript,screen]{acmart}
\AtBeginDocument{%
  \providecommand\BibTeX{{%
    \normalfont B\kern-0.5em{\scshape i\kern-0.25em b}\kern-0.8em\TeX}}}

\copyrightyear{2024}
\acmYear{2024}
\setcopyright{acmlicensed}\acmConference[CHI '24]{Workshop Pre-print: The CHI Conference on Human Factors in Computing Systems}{May 11--16, 2024}{Honolulu, HI, USA}
\usepackage{comment}
\usepackage{float}

\usepackage{xcolor,colortbl}
\definecolor{Hildaspink}{HTML}{FF99BE}

\usepackage{enumerate}  
\usepackage[shortlabels]{enumitem}

\begin{document}

\title[Sora OpenAI's Prelude]{Sora OpenAI's Prelude: Social Media Perspectives on Sora OpenAI and the Future of AI Video Generation}

\settopmatter{printacmref=false}

\author{Reza Hadi Mogavi}
% \authornote{rhadimog@uwaterloo.ca}
\email{rhadimog@uwaterloo.ca}
\orcid{0000-0002-4690-2769}

\author{Derrick Wang}
\email{dwmaru@uwaterloo.ca}
\orcid{0000-0003-3564-2532}

\author{Joseph Tu}
\email{joseph.tu@uwaterloo.ca}
\orcid{0000-0002-7703-6234}

\author{Hilda Hadan}
\email{hhadan@uwaterloo.ca}
\orcid{0000-0002-5911-1405}

\author{Sabrina A. Sgandurra}
\email{sasgandu@uwaterloo.ca}
\orcid{0000-0003-3187-263X}

% \author{Lennart E. Nacke}
% \email{lennart.nacke@acm.org}
% \orcid{https://orcid.org/0000-0003-4290-8829}

\affiliation{%
  \institution{Stratford School of Interaction Design and Business, University of Waterloo}
  \country{Canada}
}

\author{Pan Hui}
% \authornote{Pan Hui is also affiliated with the Department of Computer Science at the University of Helsinki, Finland}
\email{panhui@ust.hk}
\orcid{https://orcid.org/0000-0001-6026-1083}
\affiliation{%
  \institution{Hong Kong University of Science and Technology (Guangzhou), Hong Kong SAR and Guangzhou}
  \country{China}
}

\author{Lennart E. Nacke}
\email{lennart.nacke@acm.org}
\orcid{https://orcid.org/0000-0003-4290-8829}
\affiliation{%
  \institution{Stratford School of Interaction Design and Business, University of Waterloo}
  \country{Canada}
}

\renewcommand{\shortauthors}{Reza Hadi Mogavi et al.}

\begin{abstract}
The rapid advancement of Generative AI (Gen-AI) is transforming Human-Computer Interaction (HCI), with significant implications across various sectors. This study investigates the public's perception of Sora OpenAI, a pioneering Gen-AI video generation tool, via social media discussions on Reddit before its release. It centers on two main questions: the envisioned applications and the concerns related to Sora's integration. The analysis forecasts positive shifts in content creation, predicting that Sora will democratize video marketing and innovate game development by making video production more accessible and economical. Conversely, there are concerns about deepfakes and the potential for disinformation, underscoring the need for strategies to address disinformation and bias. This paper contributes to the Gen-AI discourse by fostering discussion on current and future capabilities, enriching the understanding of public expectations, and establishing a temporal benchmark for user anticipation. This research underscores the necessity for informed, ethical approaches to AI development and integration, ensuring that technological advancements align with societal values and user needs.
\end{abstract}

\begin{CCSXML}
<ccs2012>
   <concept>
       <concept_id>10003120.10003121</concept_id>
       <concept_desc>Human-centered computing~Human computer interaction (HCI)</concept_desc>
       <concept_significance>500</concept_significance>
       </concept>
 </ccs2012>
\end{CCSXML}

\ccsdesc[500]{Human-centered computing~Human computer interaction (HCI)}

%\hilda{I looked at other ACM generative ai papers, here are what they used}

\keywords{Generative AI (Gen-AI), Video-Generating Model, Sora OpenAI, Social Media Content Analysis.}

% \begin{figure}
%      \centering
%          \includegraphics[width=\textwidth]{content/sample.jpg}
%          \caption{A compact illustration of how the promotional gamification scheme of WB works in Stack Exchange}~\label{fig:info-graph}
% \end{figure}
\begin{teaserfigure}
\begin{minipage}[b]{0.48\textwidth}
\includegraphics[width=\textwidth]{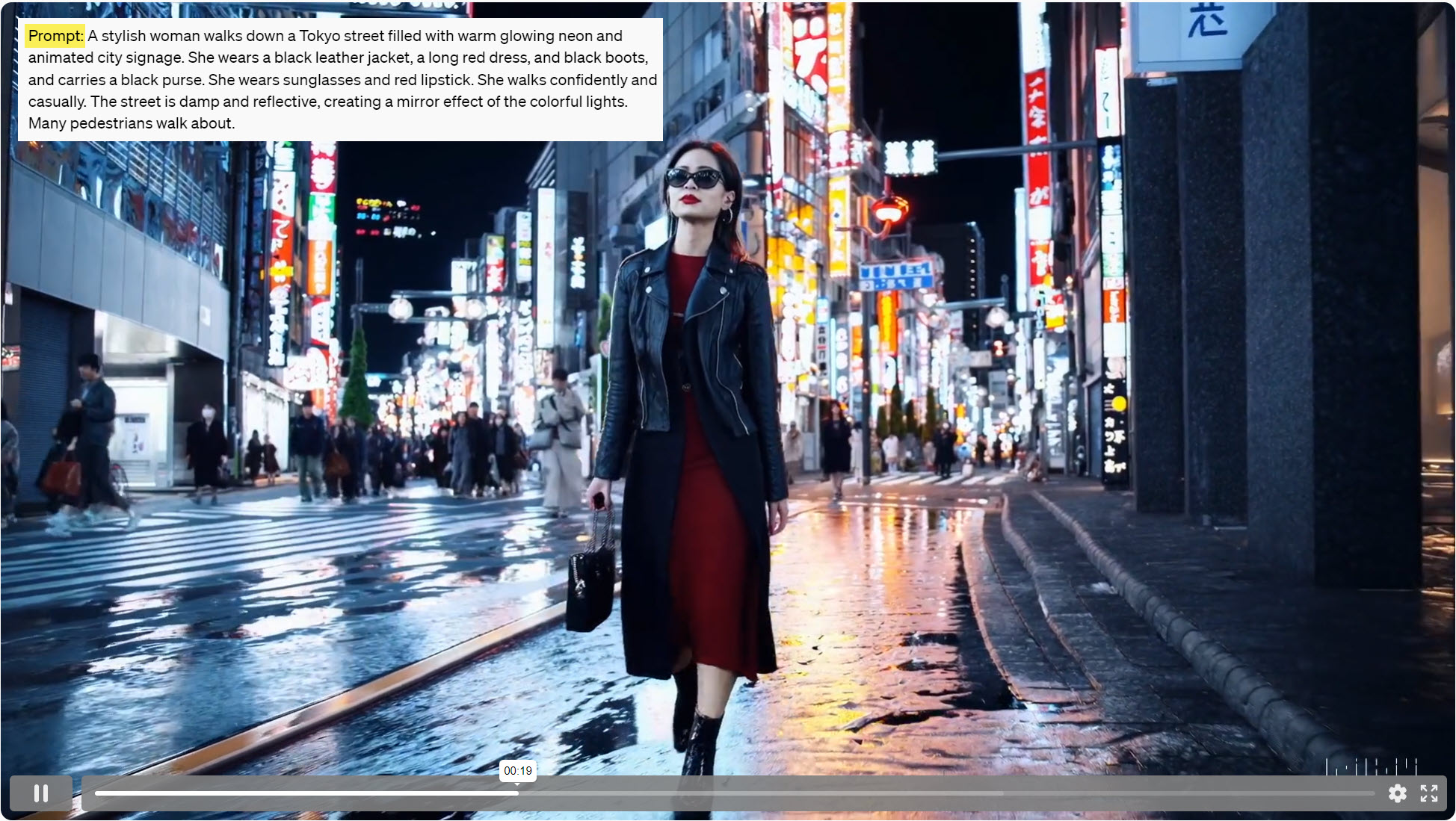}
\end{minipage}\hfill
\begin{minipage}[b]{0.48\textwidth}
\includegraphics[width=\textwidth]{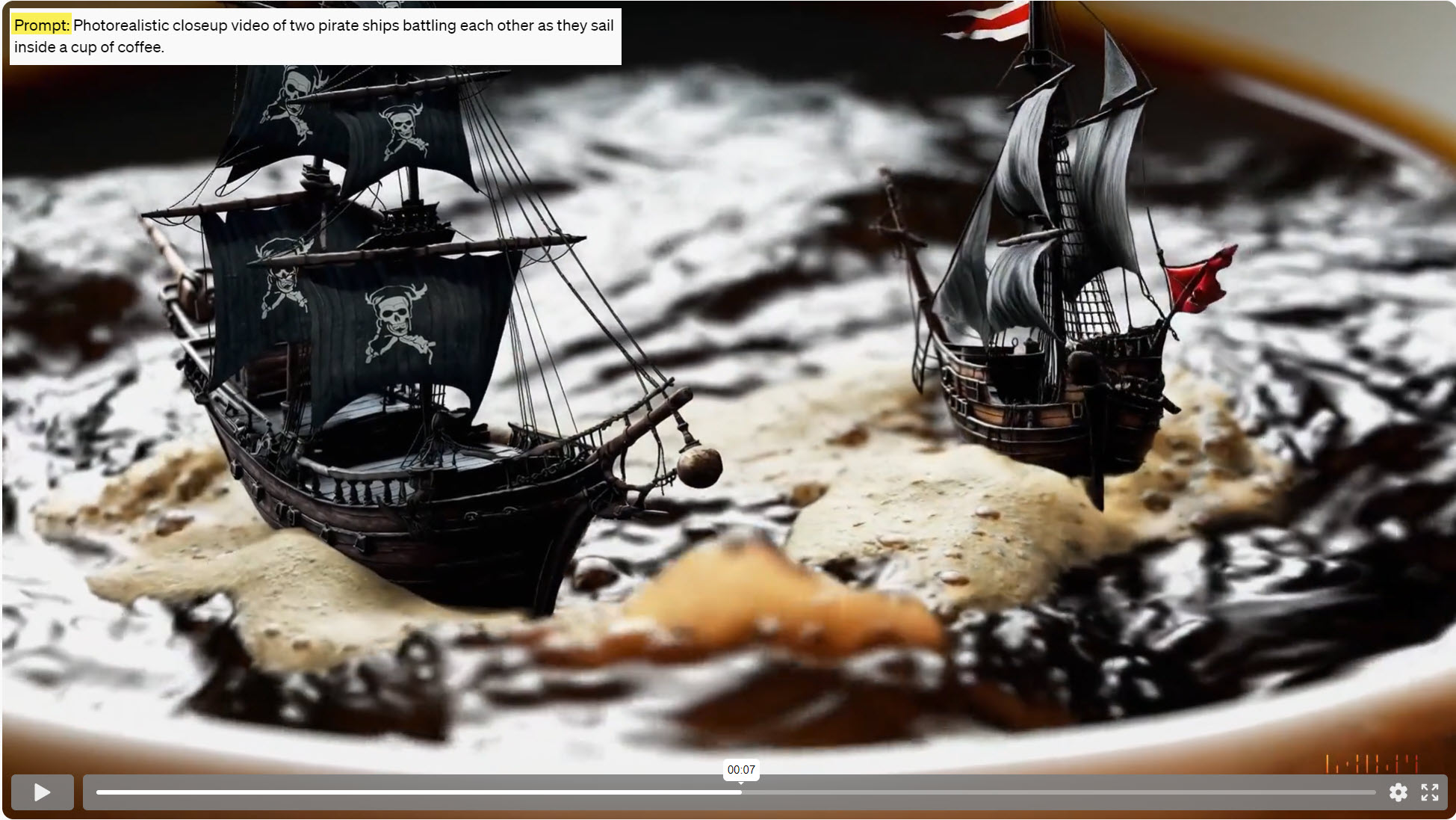}
\end{minipage}
\caption{Two snapshots from video outputs generated by OpenAI's Sora model in response to textual prompts (refer to \cite{OpenAIWebsiteref})}
\label{fig:teaser}
\end{teaserfigure}


\maketitle

\section{Introduction}\label{sec:introduction}
The remarkable advancement of Generative Artificial Intelligence (Gen-AI) has catalyzed a sweeping transformation across the domain of Human-Computer Interaction (HCI) \cite{10.1145/3544549.3573794, 10.1145/3491101.3503719, zheng2024charting}. Due to its ability to produce original material in a variety of forms, including written, visual, and auditory modes, Gen-AI has had a significant influence in fields such as art, learning, investigation, and recreation \cite{Galanter2016, zheng2024charting, zhang2023redefining, singer2022make, 10.1145/3544548.3581402}. The ubiquitous adoption of these tools by content creators has propelled Gen-AI's influence beyond its initial scope, engendering both promising prospects and formidable challenges for the HCI community \cite{zheng2024charting, HadiMogavi2024}.

Thus, it is imperative to closely monitor the emergence and advancement of novel Gen-AI technologies, given the rapid pace of progress in this field. By gaining a comprehensive understanding of human perspectives in this space (especially at an early stage \cite{HadiMogavi2024}), we can enhance aspects such as \textit{user experience}, \textit{value}, \textit{usability}, \textit{desirability}, and \textit{adoption rates} \cite{masters13}. The existing body of knowledge has demonstrated that social media platforms, specifically Reddit, can serve as a valuable tool in comprehending people's perspectives and beliefs surrounding emerging Artificial Intelligence (AI) technologies \cite{HadiMogavi2024, tian2023last}.

In this workshop paper, we will use the qualitative textual data gathered from Reddit to gain a better understanding of people's perceptions and opinions of Sora OpenAI (before it is open to the public). Sora is an advanced AI video generation tool that can create realistic and imaginative scenes based on text instructions \cite{OpenAIWebsiteref}.

At first glance, the idea of seeking people's opinions before a product's official release may seem counter-intuitive. However, this type of research enables businesses to gauge market responses, comprehend consumer expectations, and implement necessary revisions or enhancements before the public unveiling. In fact, the strategy of pre-announcing a new technology before its official debut is a deliberate tactic commonly utilized by many companies to build excitement and anticipation among potential consumers \cite{Choi2005}.

With this background in mind, our discussion in this workshop paper is guided by two research questions:
\begin{itemize}
    \item \textbf{RQ1:} What are the primary applications people envision for Sora OpenAI, and what are their expectations for its impact?
    \item \textbf{RQ2:} What concerns do people have about Sora OpenAI's integration, including potential fears and challenges?
\end{itemize}

Our qualitative data analysis has revealed that OpenAI's Sora is being viewed as a catalyst for advancement in various fields, including video marketing, game development, and educational content creation. This technological leap prompts us to reflect on the nature of creativity itself--how it may be transformed when combined with AI--and the ethical responsibilities required as we venture into this new collaborative frontier between human innovation and AI.

\textbf{Contributions.} This paper offers three key contributions to the Gen-AI and HCI communities. Firstly, it sparks an open discussion on generative AI's current capabilities, helping to pinpoint future research directions. Secondly, it deepens our understanding of public perceptions and expectations of emerging technologies, which is vital for creating user-centered designs. Lastly, acting as a temporal benchmark before the product's release, it allows for future comparisons between expected and actual outcomes, providing valuable insights into the accuracy of societal predictions. Our endeavors aspire to propel the evolution of AI technologies in a manner that is informed, ethically responsible, and meticulously aligned with user anticipations.
\section{Sora Explained}\label{sec:background}
In the quest to explore the intersection of AI and video generation, OpenAI's Sora emerges as a beacon of innovation, enabling the creation of videos from simple textual descriptions. Building upon the capabilities of previous models, Sora enhances video production with improvements in length, resolution, and aspect ratio, achieving high-definition quality for up to a minute \cite{OpenAIWebsiteref1}. To visually demonstrate Sora's prowess, Figure \ref{fig:teaser} includes snapshots from videos crafted in response to textual prompts, highlighting its adaptability and range.

While our analysis does not delve deeply into technical jargon, it is perhaps insightful to highlight some of the foundational aspects that underpin Sora's development, as outlined in OpenAI's technical report from February 15, 2024 (see \cite{OpenAIWebsiteref1}). This report connects the dots between the evolution of Large Language Models (LLMs)--known for their expansive learning from vast internet data using text tokens--and the advent of Sora. Sora innovates by applying the concept of tokens to the visual domain, using ``visual patches'' to represent and generate complex video content. This method has been previously validated for its efficiency and scalability in handling diverse visual tasks \cite{OpenAIWebsiteref1}.

Central to Sora is the use of a diffusion model, a technique that processes noisy initial inputs to produce clear, coherent visual outcomes. This is powered by transformer architecture, a framework celebrated for its scalability and effectiveness across various applications \cite{OpenAIWebsiteref1}, including language modeling \cite{ashish2017attention,brown2020language}. computer vision \cite{dosovitskiy2020image, arnab2021vivit}, and image generation \cite{chen2020generative, ramesh2021zero}. OpenAI's research highlights how these diffusion transformers are particularly adept at enhancing video model performance, demonstrating significant improvements in video quality as the model undergoes more extensive training with increased computational resources.

The introduction of Sora may mark a significant milestone in the capabilities of generative artificial intelligence, potentially opening new avenues for applications such as creative storytelling and detailed simulations of the physical world. However, alongside the technological advancements, Sora brings to light ethical considerations, particularly around the authenticity of generated content. It underscores the need for a balanced approach to innovation, where the pursuit of new possibilities is aligned with ethical standards in content creation. We look forward to discussing some of these topics with our peers at the conference.
\section{Data Collection and Qualitative Data Analysis}\label{sec:method}
Our data collection process involved meticulously gathering comments manually from the Reddit forums r/OpenAI and r/blender, which were selected for their active discussions (i.e., the highest number of posts and comments) related to Sora OpenAI. We compiled a dataset of 602 and 745 comments from these forums, respectively, including comments starting from February 16 up until February 23, 2024.

The first author used an inductive approach and Atlas.ti for the initial analysis, swiftly identifying key themes related to RQ1 and RQ2 within the Sora OpenAI Reddit discussions. This preliminary effort aimed to surface immediate insights rather than pursue exhaustive coding \cite{clarke2013successful}. The findings are intended to spark discussion at the workshop and foster future collaborations, with an understanding that a more detailed coding effort will follow.
\section{Summary of Our Main Findings}\label{sec:findings}
Our initial analysis of social media content has uncovered notable insights regarding the public's anticipation and concerns surrounding Sora OpenAI.

\textbf{RQ1: Envisioned Applications and Impact.}
The public foresees Sora OpenAI as a revolutionary force in content creation, with particular benefits for small businesses and video marketers in the short video era. Users predict that Sora will democratize video marketing by enabling the animation of product photos and the integration of dynamic storytelling. This capability is expected to make video production more accessible and economical, as one Reddit user explains: ``\textit{The prospect of dramatically lowering the cost and complexity of building dynamic 3D worlds is a dream to me. It means that small teams and inspired artists can build the thing they have in their heads.}''

Within the gaming industry, Sora is anticipated to expedite asset creation and cinematics, aiding both indie developers and major studios. Further, Sora's potential extends into personalized entertainment, where viewers could see themselves as the main character in films, a prospect that one Reddit user believes could become normal in five years.

Significantly, the community also envisions Sora enhancing educational content. The AI's capabilities to generate illustrative videos could offer students immersive experiences, such as visualizing historical events or understanding complex scientific concepts. A Reddit user highlights this potential application: ``\textit{You can also use it for other useful cases like educational videos like physics videos or a nice illustrative videos on war or a rendition of how Egypt might have looked like for students to take a look at without wasting too much effort.}''

Moreover, there is a sentiment that, as Sora continues to evolve, it will improve its ability to interpret users' creative visions, potentially ushering in a new era of digital storytelling. A Reddit user expresses this sentiment: ``\textit{IMO the further we go down this road, the better the models will get at interpreting the creative vision of the user. If we have powerful tools that can bring regular people's creativity and storytelling to life, a whole new wave of artists will be born.}'' They continue, asserting the importance of individual creativity: ``\textit{AI generated `stories' and `jokes' have never been all that great, but every human has the ability to tell incredible stories, and now they'll have the technology to conceptualize those stories in their own unique and creative way.}''

\textbf{RQ2: Integration Concerns and Challenges.}
While the potential of Sora OpenAI is met with excitement for its applications, there are significant concerns regarding its societal integration. The thematic analysis surfaces critical apprehensions from various perspectives.

The threat to creative jobs is a pressing concern, with fears that AI advancements could make human artistry non-competitive. A Reddit user voices distress about the future of creative education: ``\textit{As a high school senior going into 3D modeling for college, I'm absolutely terrified that my education might be worthless in a matter of years.}'' This sentiment is echoed by another user, who fears that the creative industry might cease to exist, affecting not only artists, but other industries as well.

The public is also concerned about the ethical implications of AI training methods and the cultural representation in AI-generated content. Criticism pointed out that AI may perpetuate existing biases and misrepresentations, as a Reddit user explains: ``\textit{Since it can only take from an already existing database of images as its reference, it will take in all of the internet without any ethical filter, with all its good and bad parts. i.e racism, sexism, difficult to find or recreate POC features etc. For example, it renders most Indians with eurocentric features because of lack of south Asian references in digital painting online.}''

Some individuals criticize the attitudes surrounding AI and creativity, suggesting that the essence of art is being misunderstood and undervalued. One user laments the current discourse: ``\textit{I'm just pissed about how all these AI bros who have never been creative in their lives are degrading the conversation around art and creativity with their inane ignorant opinions.}'' They continue to defend traditional artistic practices: ``\textit{But these people seem to have the attitude that the creative process is a problem to be overcome rather than the entire point.}''

Another technical concern is the unpredictable nature of AI in creative workflows. A user points out the limitations of AI as an artistic tool, stating, ``\textit{Imagine a prompt engineer trying to fit in an art pipeline, he would simply have to depend on pure luck and it would be almost impossible to recreate the art as the `prompt engineer' himself doesn't know the workings behind it.}''

The diverse concerns expressed indicate the pressing need for a systematic approach to address the complex issues of disinformation, bias, and ethical challenges surrounding AI use and societal integration. Our investigation shows a clear demand for initiatives that promote media literacy, fact verification, ethical frameworks, and regulatory protocols for overseeing AI-generated content, including tools like Sora. We believe that collaboration among all stakeholders is crucial for the development, deployment, and use of AI. This ensures a balance between innovation and ethical standards, while also protecting human creativity's integrity.

% put other sections as files in the content folder...

% \begin{acks}
% Thanks for all the fish.
% \end{acks}

\bibliographystyle{ACM-Reference-Format}
\bibliography{sample-base}

% \appendix









\end{document}